\documentclass[journal]{IEEEtran} 
\usepackage[numbers]{natbib}
\usepackage[english]{babel}
\ifCLASSINFOpdf
  \usepackage[pdftex]{graphicx}
\else
\fi
\usepackage{siunitx}
\usepackage{xfrac}
\usepackage[cmex10]{amsmath}
\usepackage{mathtools}
\usepackage[dvipsnames]{xcolor}

\usepackage{booktabs}
\usepackage{multirow}
\usepackage{url}
\usepackage{color} 
\begin{document}
\title{Fluke 8588A DMM Sampling Performance}

\author{Rado~Lapuh, Jan Ku\v{c}era, Jakub Kov\'{a}\v{c}, Bo\v{s}tjan Volj\v{c}
\thanks{R. Lapuh is with the Metrology Institute of the Republic of Slovenia, 
3000 Celje, Slovenia, (e-mail: rado.lapuh@gov.si). 
Jan Ku\v{c}era and Jakub Kov\'{a}\v{c} are with Czech Metrology Institute (CMI), Okru\v{z}ni 31, 638 00 Brno, Czech Republic
B. Volj\v{c} is with Slovenian Institute of Quality and Metrology,
}
}

\markboth{}
{Shell \MakeLowercase{\textit{et al.}}: Bare Demo of IEEEtran.cls for Journals}
\maketitle

\begin{abstract}
This paper reports on an evaluation of the Fluke 8588A digital multimeter (DMM) sampling performance and comparison to Keysight 3458A DMM sampling performance.
The design of 8588A type DMM shows both similarities and striking differences in design compared to 3458A type DMM.
Apart from their specified sampling capabilities, measurements were performed to evaluate frequency response, noise performance, including a 1\hspace{0.4mm}/\hspace{0.4mm}\textit{f} type noise and a timing jitter induced noise, as well as phase measurement performance. \end{abstract}


\IEEEpeerreviewmaketitle

\section{Introduction}
\IEEEPARstart{S}{oon} after the 3458A DMM was introduced to a market in 1988, metrologists in the National Metrology Institutes (NMI) started using its novel sampling capabilities for various measurements, achieving groundbreaking performance increase, particularly in terms of lowering measurement uncertainties \cite{Swerlein1991,Cage1991,Svensson1996,Ramm1999,Ihlenfeld2001,Mohns2009,Iuzzolino2007} in various measurement areas. 
This was achieved by using a custom made multi-slope integrating analogue-to-digital converter (IADC), which exhibits a few key properties built in by design.
First, the IADC has a variable resolution, which is a function of a programmable aperture time.
Second, the aperture time provides for a calculable filter, which effectively attenuates high frequency (interference) signals before they can interfere with the resulting quantisation process.
Third, this filter function can be equalised almost exactly even for a large attenuation values, achieving a correction ratios of up to \num{5000}:\num{1} \cite{Swerlein1991}.
This resulted in a state of the art sampler for low frequency signals \cite{Espel2009}.

The IADC technique has since evolved into a wildly popular delta-sigma conversion technique \cite{Horowitz_2015}.
The NI-5922 flexible resolution digitiser became popular in many NMIs for performing state of the art measurements and it arguably offers the top performance that delta-sigma conversion technique can deliver. 
However, its performance never challenged the 3458A performance at low frequency sampling applications.
After three decades, the Fluke 8588A multimeter (DMM) appears to be the first DMM in almost thirty years to challenge the state of the art Keysight 3458A DMM sampling performance.
The first evaluation of its sampling performance with comparison to 3458A is presented in this paper.
The discussion is limited solely for sampling voltage signals.
As this evaluation only deals with sampling voltage signals, it can in no case prejudice complete performance of any instrument discussed.

A precision \v{C}MI SWG03 DDS source \cite{Kucera2020} and Keysight 33622A arbitrary waveform synthesizer \cite{Keysight33622A_OperatingManual} were used for all measurements requiring a sine wave input signal.

\section{ADC technology}\label{sec:ADC performance}

\subsection{DMM ADC}\label{sec:DMM ADC}
DMMs are primarily intended to accurately measure voltage, current, resistance and possibly other quantities. 
To achieve this, the ADC used must be stable, linear and have a high resolving resolution, while the high conversion speed is not required. 
The ADC technology of choice is an integrating or charge balance ADC (IADC), dominating the DMM market for decades.
Its multi-slope version, developed for then Hewlett-Packard 3458A provided for an unprecedented performance increase in stability, linearity and resolution. 
Importantly, this was not sacrificed by an excessively slow performance.
On the contrary, also the maximum sampling rate was increased from a solid \SI{1350}{\hertz} in previous model (HP 3457A) \cite{HP3457A_OperatingManual} to a then whooping \SI{100}{\kilo\hertz} for an IADC while still maintaining \SI{16}{bit} resolution \cite{HP_Journal}.
To cope with even higher frequency repetitive signals, the sample-hold circuitry is employed to freeze a signal until it is converted by its IADC.
This circuitry allows for only \SI{50}{\kilo\hertz} sampling frequency, but a short aperture time of \SI{2}{\nano\second} allows for a sub-sampling scheme, technically employing \SI{100}{\mega\hertz} sampling rate and capturing repetitive signals with frequencies approaching \SI{50}{\mega\hertz}.
However, this sample-hold measurement path is far less accurate than a direct DCV path.

8588A type DMM introduces a slightly different concept, using IADC (Fluke calls this type of ADC a Charge Balancing ADC or CB) for slow sampling rates up to \SI{4,7619}{\kilo\hertz} and an \SI{18}{bit} successive approximation (SAR) ADC for fast sampling rates up to \SI{5}{\mega\hertz}.
Additionally, as a new product, 8588A offers much larger internal memory (15 M samples) and a set of interfaces for transferring commands and data (GPIB, USB and Ethernet).
The CB and SAR technologies are said to work together\cite{Fluke_ADC_technology} for the best precision and best digitisation, but it is not disclosed neither how this is implemented nor what are the actual benefits, except of actually covering combined sampling rates from \SI{0,1}{\hertz} up to \SI{5}{\mega\hertz}.
8588A actually provides for three modes of operation\cite{Fluke_ADC_technology,Fluke_8588A_OperatorsManual}:
\begin{itemize}
  \item CB mode can be used for aperture times from \SI{100}{\micro\second} up to \SI{10}{\second}, with minimum trigger interval equal to aperture time plus \SI{170}{\micro\second}. 
  \SI{25}{\micro\second} is used for balancing zero before the measurement aperture and \SI{85}{\micro\second} is used afterwards for sub-reference ramps of the multi-slope operation. 
  There seems to be an additional \SI{60}{\micro\second} necessary for the internal processing before a new valid trigger can be accepted. 
  The CB mode is the most stable and the most accurate mode of operation\cite{Fluke_8588A_ProductSpecifications}. 
  
  \item SAR mode can be used for aperture times below \SI{100}{\micro\second} down to "zero" aperture time. 
  Here the aperture is emulated by averaging the required number of samples from SAR ADC, which is internally sampling at \SI{5}{\mega\hertz} sample rate. 
  Additional \SI{30}{\micro\second} is required at each conversion, yielding a highest sampling frequency of \SI{33.333}{\kilo\hertz}. 
  The use of this additional \SI{30}{\micro\second} time is not disclosed.
  
  \item DIGITIZE (DIG) mode employs the SAR DAC directly, without any additional delays, enabling the full \SI{5}{\mega\hertz} sampling speed. Aperture time is still programmable from \SI{0}{\nano\second} to \SI{3}{\milli\second} and it uses the same technique as in SAR mode to emulate the analogue CB aperture time function.
  Manufacturer voltage accuracy specifications\cite{Fluke_8588A_ProductSpecifications} for SAR mode and Digitizing mode are virtually the same.
  
\end{itemize}
\subsection{DMM sampling resolution}\label{sec:DMM sampling resolution}
\begin{figure}
\center
\includegraphics[scale = 1] {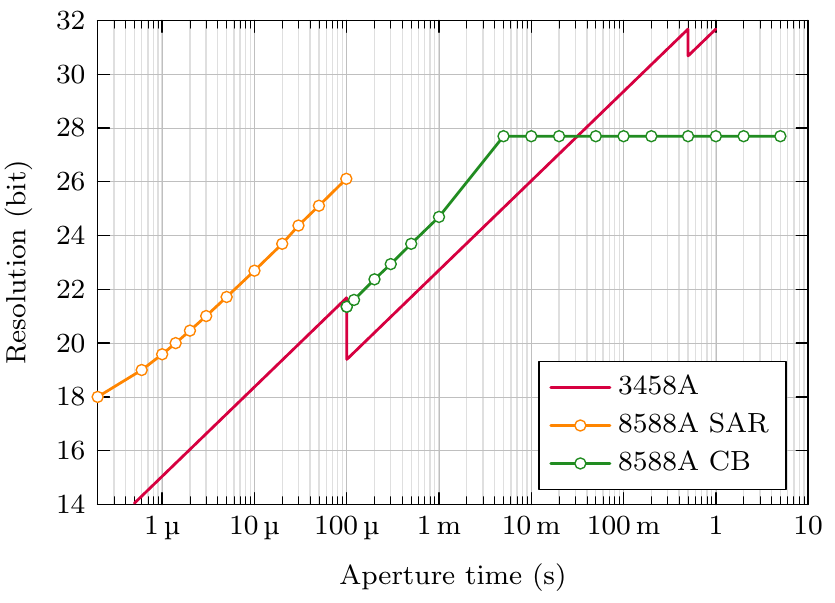}
\caption{DMM analogue to digital converter resolving resolution in bits for nominal Range input level. Circles denote points in which this resolution was extracted form sampled data. For 3458A, this was achieved on more points, which are not indicated on this graph.}
\label{fig:8588A_resolution_bits}
\end{figure}

The sampling resolution discussed in this report refers to the ADC quantisation resolution and not to the usual DMM resolution specification, which also includes other imperfections like differential non-linearity and noise\cite{HP_Journal_Goecke}.
The sampling resolution can be easily determined by reading a scale factor which is used to convert binary values back to voltage values\cite{LapuhBook}. 
This was used to retrieve the 3458A sampling resolution, but was not a useful approach for 8588A as this function was not working on the available pair of 8588A. 
Alternatively, a number of samples were taken with input terminals shorted and the record was sorted from the lowest to the highest value measured. 
The minimum step was easily retrieved, indicating the ADC numerical resolution at a number of aperture times selected.
The retrieved sampling resolution for both DMMs is shown on figure \ref{fig:8588A_resolution_bits}.
The SAR sampling resolution increases with the increasing aperture time as the averaging function provides for the otherwise fixed SAR ADC resolution increase.
Further, 8588A type DMM provides for a higher sampling resolution than 3458A type DMM from \SI{100}{\micro\second} to \SI{30}{\milli\second} aperture time.
However, from \SI{4}{\milli\second} aperture time and longer, the 8588A limits the CB ADC resolution at \SI{28}{bits}, while 3458A limit is imposed at \SI{32}{bits}.
Again, the sampling resolution described here is not providing any meaningful DMM performance parameter, but it provides an upper limit of internal ADC resolving power and perhaps a glimpse on its internal operation.
\section{DMM noise performance}\label{sec:DMM noise performance}
There is a difference in the 8588A DMM design that has to be noted here to better understand the performance being reported later on. 
Its CB mode signal path \SI{28}{\kilo\hertz} bandwidth is much lower than the 3458A's DCV \SI{150}{\kilo\hertz} bandwidth.
This would, in principle, reduce the overall noise of the DMM, with the price paid on the limited frequency response linearity at low frequencies.
Further, 8588A employs chopper amp stabilisation to achieve superior offset performance \cite{Fluke_ADC_technology}.
This would push the pink noise corner frequency to much lower values, but could potentially degrade the overall signal to noise ratio.

\subsection{Signal to Noise Ratio}\label{sec:Signal to Noise Ratio}
The signal to noise ratio was measured by short circuiting the input terminals and performing sampling at different aperture times.
\num{128} samples were taken in a sample burst and three bursts were taken to further average the result at each aperture time. 
Sampling frequencies were chosen so as to cancel the mains signal frequency interference. 
From the burst samples, the noise voltage $\sigma_n$ was estimated as a median value of the FFT spectrum noise amplitude, thus rejecting interference signals originating from the mains frequency signal or other sources producing steady frequency signals.
The $\mathit{SNR}$ was finally calculated as a ratio of nominal range RMS voltage $R / \sqrt{2}$ to the noise voltage
\begin{equation}
\mathit{SNR}=\frac{\mathrm{R}}{\sigma_n \sqrt{2}},
\label{eq:SNR}
\end{equation}

\begin{figure}
\center
\includegraphics[scale = 1] {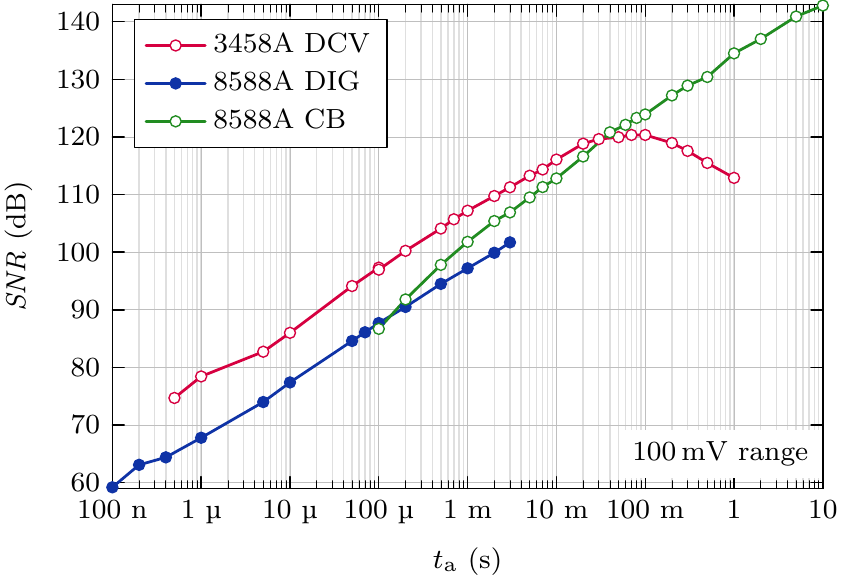}
\caption{\SI{100}{\milli\volt} range signal to noise ratio.}
\label{fig:SNR 100mV}
\end{figure}

\begin{figure}
\center
\includegraphics[scale = 1] {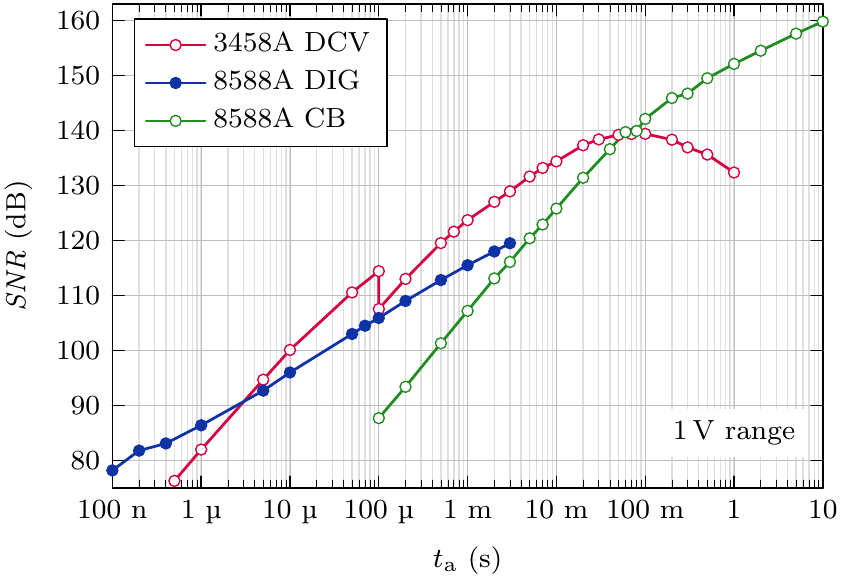}
\caption{\SI{1}{\volt} range signal to noise ratio.}
\label{fig:SNR 1V}
\end{figure}

\begin{figure}
\center
\includegraphics[scale = 1] {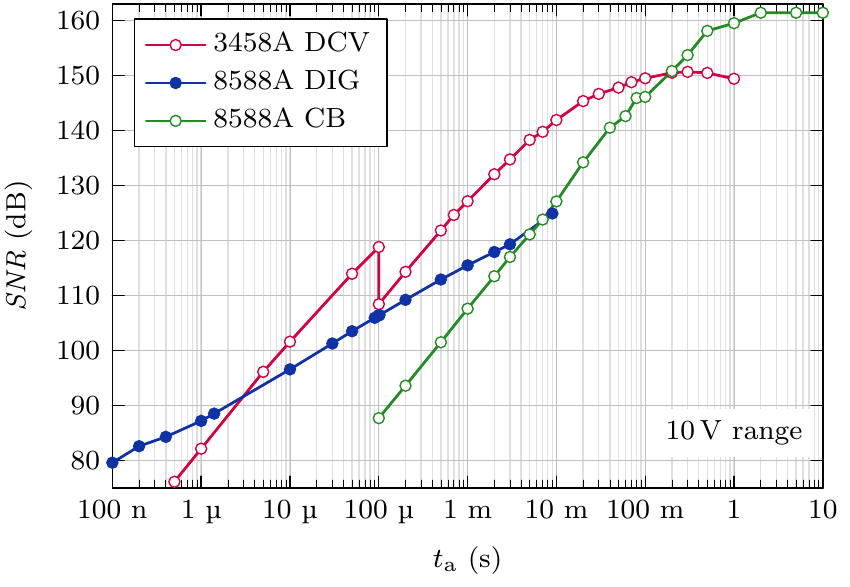}
\caption{\SI{10}{\volt} range signal to noise ratio.}
\label{fig:SNR 10V}
\end{figure}

\begin{figure}
\center
\includegraphics[scale = 1] {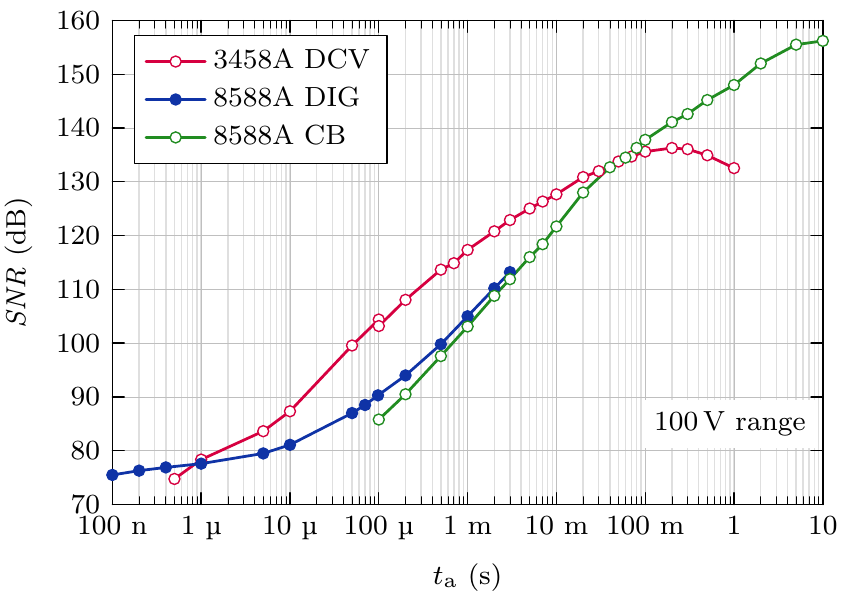}
\caption{\SI{100}{\volt} range signal to noise ratio.}
\label{fig:SNR 100V}
\end{figure}

\begin{figure}
\center
\includegraphics[scale = 1] {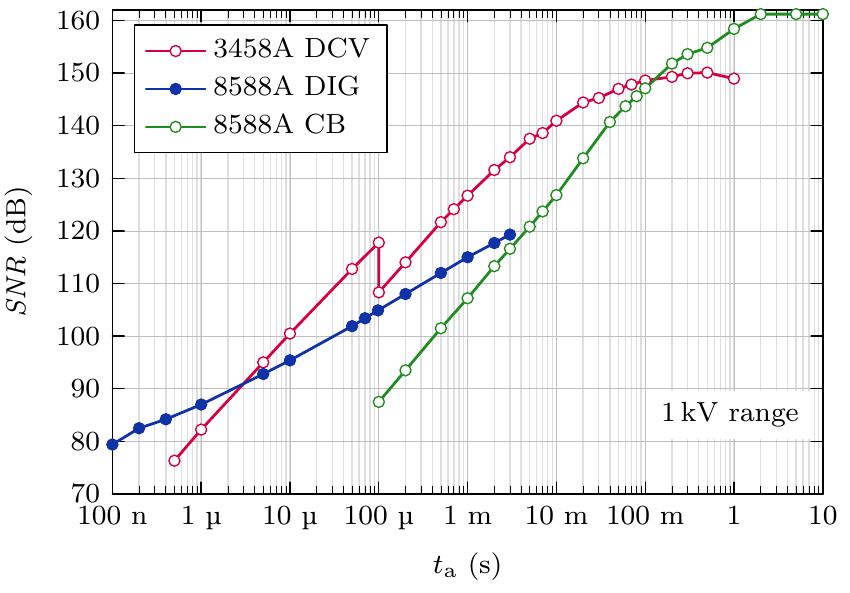}
\caption{\SI{1}{\kilo\volt} range signal to noise ratio.}
\label{fig:SNR 1kV}
\end{figure}
Results show a stark performance difference in signal to noise ratio between 3458A (DCV mode) \cite{Lapuh2016} and 8588A (CB mode) DMMs.
At aperture times from \SI{100}{\micro\second} up to \SI{10}{\milli\second}, the 3458A $\mathit{SNR}$ is larger from \SI{20}{\decibel} to \SI{15}{\decibel} on \SI{10}{\volt} and \SI{1}{\kilo\volt} ranges.
This difference is smaller on other ranges where the input stage amplifiers plays a more dominant role.
From aperture time at around \SI{100}{\milli\second} and up, the $\mathit{SNR}$ of 3458A stops increasing and actually starts to decrease due to increasing $1/f$ noise \cite{LapuhBook}, while the $\mathit{SNR}$ of 8588A increases further up to \SI{161.4}{\decibel}, being actually limited by the \SI{28}{bit} resolution limit of the CB analogue to digital converter.
The 8588A DIG $\mathit{SNR}$ performance, which was found to be virtually identical to its SAR performance at aperture times below \SI{100}{\micro\second}, shows a slower increase with increasing aperture time.
Nevertheless, the $\mathit{SNR}$ in DIG mode is better than in CB mode for all aperture times available for DIG mode (up to \SI{3}{\milli\second}, except on \SI{100}{\milli\volt} range, where the input amplifier reduces the $\mathit{SNR}$ in DIG mode by \SI{20}{\decibel} across all aperture times. 

\subsection{$1/f$ Noise}\label{sec:Pink Noise}
$1/f$ noise was measured in such a way that a long continuous sampling bursts were captured in a record and an FFT was used to estimate the noise spectral density frequency response \cite{Heinzel2002}. 
Number of samples in a record was chosen so as to effectively cover the lowest frequency of interest with a fixed sampling time of \SI{9}{\milli\second}. 
For 3458A, this frequency was \SI{3}{\milli\hertz} and for 8588A in CB mode, this frequency was \SI{100}{\micro\hertz}.
To reach this low frequency result, the record length of up to \num{1000000} samples was required, which resulted in continuous sampling lasting up to \SI{2.5}{\hour} for a single record.
Internal timer was used to dictate the sampling time.

\begin{figure}
\center
\includegraphics[scale = 1] {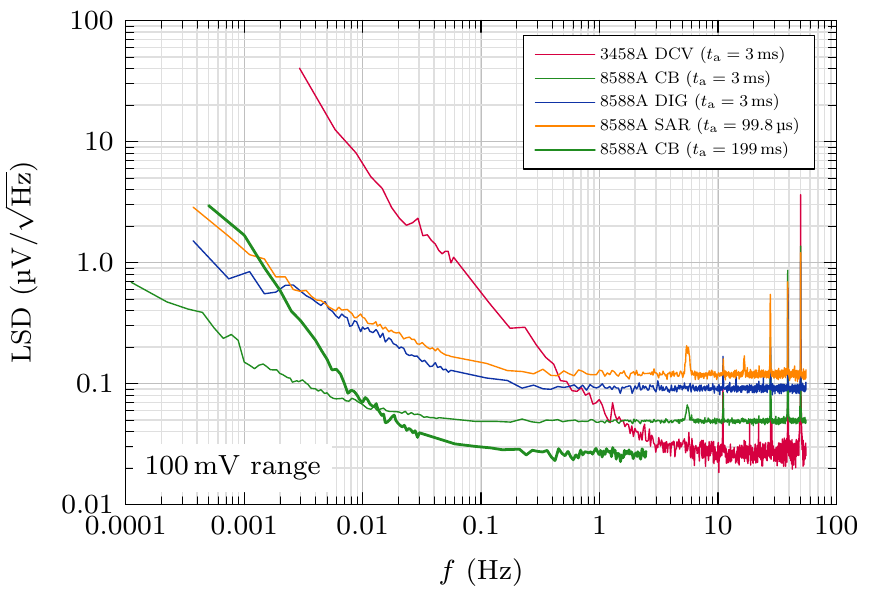}
\caption{\SI{100}{\milli\volt} range $1/f$ noise spectral density.}
\label{fig:PinkNoise 100mV}
\end{figure}

\begin{figure}
\center
\includegraphics[scale = 1] {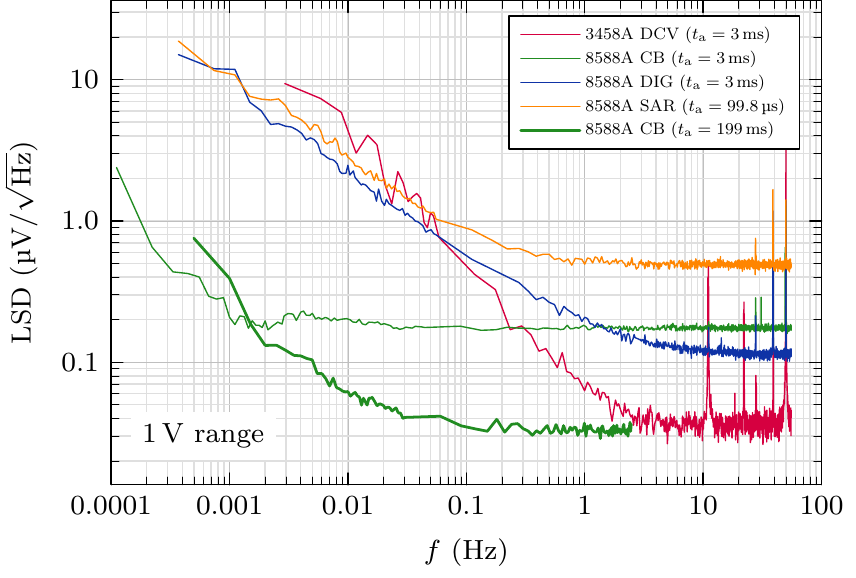}
\caption{\SI{1}{\volt} range $1/f$ noise spectral density.}
\label{fig:PinkNoise 1V}
\end{figure}

\begin{figure}
\center
\includegraphics[scale = 1] {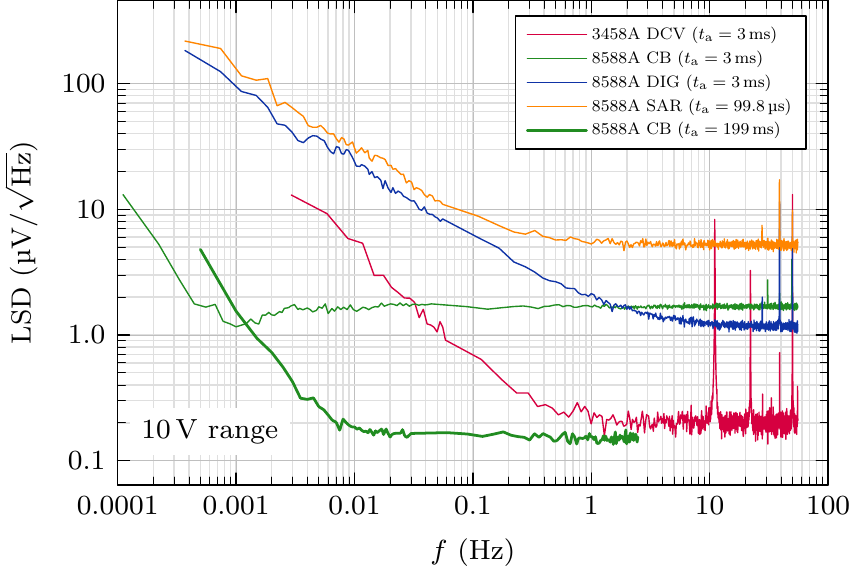}
\caption{\SI{10}{\volt} range $1/f$ noise spectral density.}
\label{fig:PinkNoise 10V}
\end{figure}

\begin{figure}
\center
\includegraphics[scale = 1] {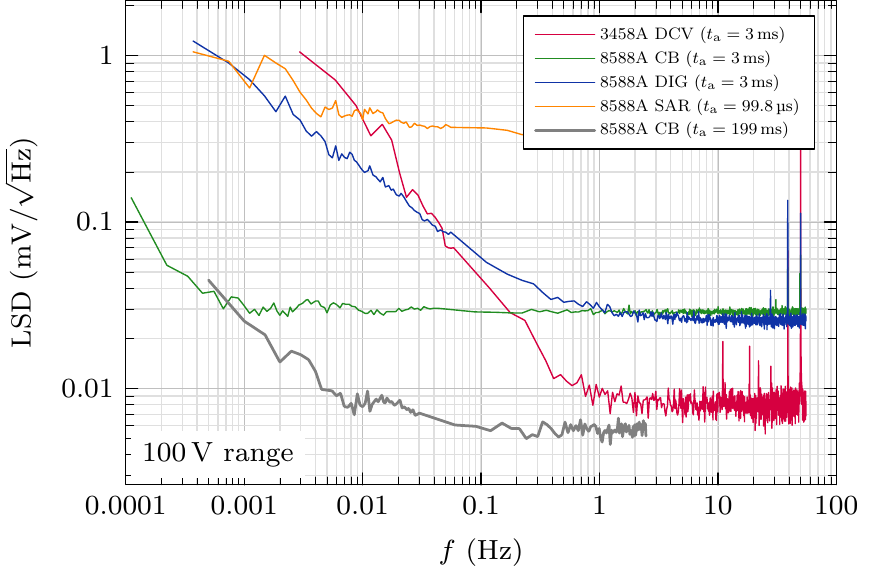}
\caption{\SI{100}{\volt} range $1/f$ noise spectral density.}
\label{fig:PinkNoise 100V}
\end{figure}

\begin{figure}
\center
\includegraphics[scale = 1] {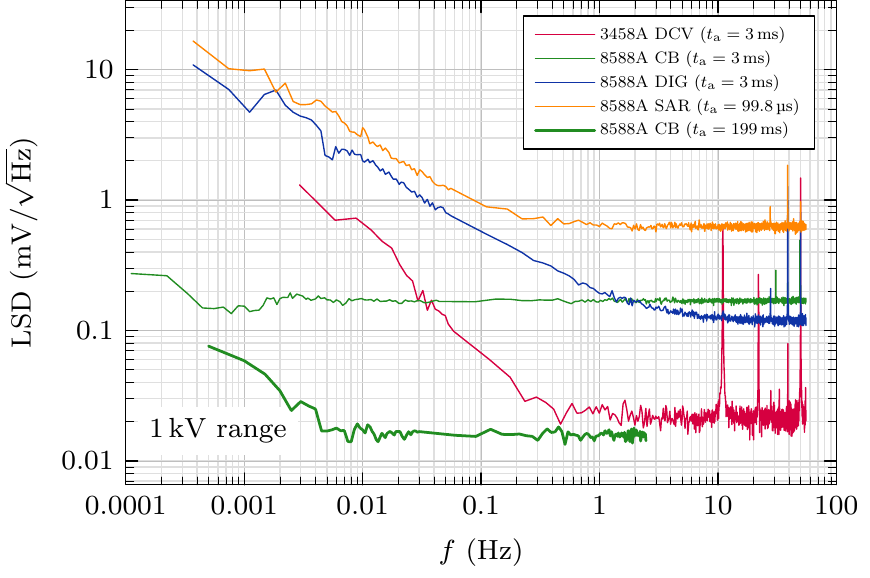}
\caption{\SI{1}{\kilo\volt} range $1/f$ noise spectral density.}
\label{fig:PinkNoise 1kV}
\end{figure}

A single non-windowed FFT results in a fairly noisy linear spectral density.
That noise can be smoothed out by averaging FFT results from more sample bursts, but in this case where a single sample burst takes minutes or even hours, this technique becomes even more time consuming.
Instead, a single record was used for averaging at higher frequencies by splitting the record into more equal length sub-records, performing FFT and averaging the results. 
The original record was split into two, three, four and so on equal length sub-records and the lowest frequency FFT result, not influenced by a residual DC level leakage (this typically being a second or a third FFT bin) was recorded as a final result. This way, the resulting linear spectral density was gradually smoothed toward higher frequencies.
Results are shown on Figures \ref{fig:PinkNoise 100mV} to \ref{fig:PinkNoise 1kV}.
They are collected together for each of five DMM ranges and cover five responses each:
\begin{itemize}
    \item 3458A IADC response at \SI{3}{\milli\second} aperture time and covering frequencies from \SI{3}{\milli\hertz} to \SI{55.6}{\hertz}.
    \item 8588A CB response at \SI{3}{\milli\second} aperture time and covering frequencies from \SI{100}{\micro\hertz} to \SI{55.6}{\hertz}. 
    \item 8588A CB response at \SI{3}{\milli\second} aperture time and covering frequencies from \SI{500}{\micro\hertz} to \SI{2.47}{\hertz}.
    \item 8588A SAR response at \SI{99.8}{\micro\second} aperture time and covering frequencies from \SI{370}{\micro\hertz} to \SI{55.6}{\hertz}.
    \item 8588A DIG response at \SI{3}{\milli\second} aperture time and covering frequencies from \SI{370}{\micro\hertz} to \SI{55.6}{\hertz}.
\end{itemize}
Frequency spikes between \SI{10}{\hertz} and \SI{55.6}{\hertz} are a byproduct of mains voltage frequency components and possibly other internal DMM or external sources and can be ignored in terms of pink noise analysis.
8588A clearly has a stronger white noise component compared to 3458A at \SI{3}{\milli\second} aperture time, supporting the measurements and discussion from the previous chapter.
However, 8588A excels in $1/f$ corner frequency, which is pushed down to \SI{10}{\milli\hertz} or so on \SI{100}{\milli\volt} range, and further down to \SI{1}{\milli\hertz} or so on all other ranges.
It appears that the 8588A SAR mode is much more noisy than 8588A DIG mode, presumably due to CB zero operation that precedes each measurement in this mode and adds its own noise to the final result.
The $1/f$ corner frequency for 8588A DIG and 8588A SAR modes seems to be one order of magnitude or so lower then $1/f$ corner frequency for 3458A.
Howevert, here it is more important to look on the actual level of linear spectral density, as the initial white noise for any 8588A mode is already much higher and is thus masking any earlier $1/f$ noise to appear as dominant noise component. 
This in effect moves the $1/f$ corner frequency to a lower value, so the corner frequency itself is not a complete $1/f$ noise performance indicator and should be evaluated together with the white noise level at higher frequencies.

\subsection{Jitter Noise}\label{sec:Jitter Noise}
Jitter noise appears in a sampled signal as a consequence of sample timing variations \cite{Lapuh2015a}.
A signal to noise ratio due to a random sampling time jitter is given by
\begin{equation}
    \mathit{SNR_\mathrm{j}}= -20 \cdot log_{10}(2 \pi f t_\mathrm{j}),
    \label{eq:SNRjitter}
\end{equation}
where $f$ is signal frequency and $t_\mathrm{j}$ is a standard deviation of the time jitter.
$\mathit{SNR_\mathrm{j}}$ defined here is referenced to the signal amplitude applied, from where it is clear that the jitter noise is not existent if there is no signal applied.
From the equation \eqref{eq:SNRjitter} it follows that for increasing signal frequency the $\mathit{SNR_\mathrm{j}}$ will drop by \SI{20}{\decibel} per decade.
The combined $\mathit{SNR_\mathrm{c}}$ is given by
\begin{equation}
    \mathit{SNR_\mathrm{c}}=-20 \cdot log_{10}\sqrt{\left( \frac{\sigma_\mathrm{n}}{V} \right)^2+\left(2 \pi f t_\mathrm{j}\right)^2},
    \label{eq:SNRcombined}
\end{equation}
where $\sigma_\mathrm{n}$ is a white noise RMS voltage.
$\sigma_\mathrm{n}$ limits the highest achievable $\mathit{SNR_\mathrm{c}}$, which starts to roll down as the $\mathit{SNR_\mathrm{j}}$ becomes the dominant $\mathit{SNR_\mathrm{c}}$ component.

\begin{figure}
\center
\includegraphics[scale = 1] {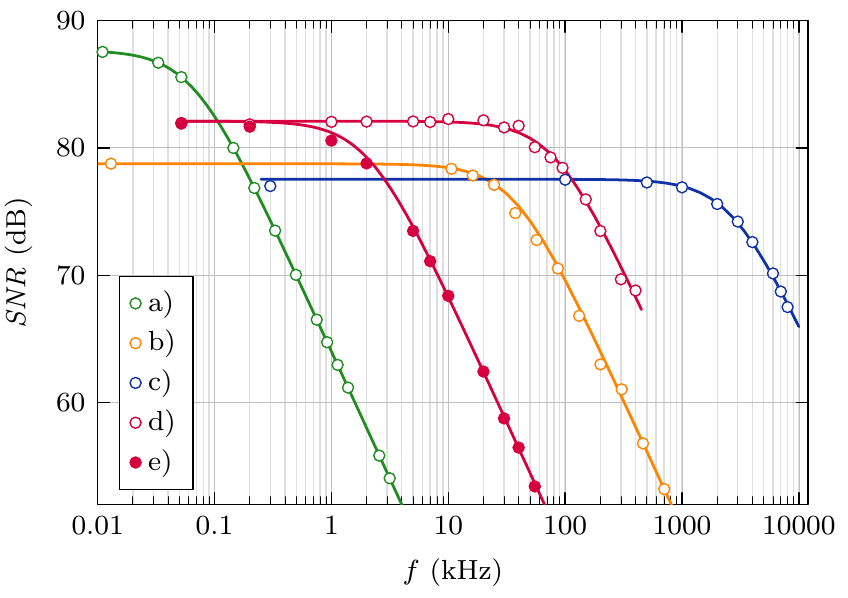}
\caption{8588A DMM charge balance ADC jitter noise induced signal to noise ratio roll-off.
a) 8588A in CB mode (sample: TIMER), 
b) 8588A in SAR mode (sample: TIMER), 
c) 8588A in DIG mode (sample: TIMER), 
d) 3458A in DCV mode (sample: TIMER), 
e) 3458A in DCV mode (sample: EXT).
3458A in DCV mode using a Master Slave synchronisation is not plotted as it is virtually identical to e).}
\label{fig:Jitter noise CB}
\end{figure}

Measurements were performed by using fixed DMM sample parameters in each mode and sampling full scale input sine wave signal at different frequencies.
The shortest available aperture times were used so that aperture roll-off would not affect measurement results.
It is further important that the sine wave amplitude remains flat within approximately \SI{+- 0.5}{\deci\bel} over the whole tested frequency range.
Additionally, it is important that the source $\mathit{SNR}$ at highest frequencies remains higher then the measured $\mathit{SNR}$ as estimated from the sampled values.
That would assure that the DMM jitter and not the source jitter is measured.
A Keysight 33622A arbitrary waveform synthesizer with amplitude flatness of \SI{+- 0.25}{\deci\bel} for frequencies up to \SI{60}{\mega\hertz} and specified time jitter of \SI{2}{\pico\second} would meet those requirements.
Measurement results plotted on Fig. \ref{fig:Jitter noise CB} clearly show behaviour described by \eqref{eq:SNRcombined}, where the results for 3458A are taken from \cite{Lapuh2015a}.
The time jitter values shown in Table \ref{tab:JitterNoise} were deduced from this measurement results using Equation \ref{eq:SNRcombined}.
\begin{table}
  \caption{Estimated effective jitter noise values from results plotted on Fig. \ref{fig:Jitter noise CB}. $t_\textrm{s}$ and $t_\textrm{a}$ are sampling parameters used for each DMM and Mode combination measurements.}
  \label{tab:JitterNoise}
  \begin{center}
  \begin{tabular}{cccccc}
  \toprule
  DMM    & Mode & Timebase used   & $t_\textrm{s}$ & $t_\textrm{a}$ & $t_\mathrm{j}$         \\
  \midrule 
  3458A & DCV & Timer             & \SI{10}{\micro\second} & \SI{1}{\micro\second} & \SI{145}{\pico\second} \\
  3458A	& DCV & External          & \SI{10}{\micro\second} & \SI{1}{\micro\second} & \SI{6}{\nano\second}   \\
  3458A	& DCV & Master-Slave      & \SI{10}{\micro\second} & \SI{1}{\micro\second} & \SI{6.8}{\nano\second} \\
  8588A	& CB  & Timer or External & \SI{20}{\micro\second} & \SI{110}{\micro\second} & \SI{100}{\nano\second} \\
  8588A	& SAR & Timer or External & \SI{30}{\micro\second} & \SI{0}{\second} & \SI{493}{\pico\second} \\
  8588A	& DIG & Timer or External & \SI{200}{\nano\second} & \SI{0}{\second} & \SI{<=7.7}{\pico\second} \\
  \bottomrule
  \end{tabular}
  \end{center}
\end{table}
For the 8588A in DIG mode result and lacking any performance test on the source, it is not completely clear if the DMM or the Keysight 33622A was the source of $\mathit{SNR}$ roll off.
Nevertheless, the results safely conclude that the 8588A time jitter in DIG mode is \SI{<=7.7}{\pico\second}.

\section{DMM amplitude response}\label{sec:DMM amplitude response}
\subsection{Aperture time roll-off}\label{sec:Aperture time roll-off}
The integrating analogue to digital converter (or sometimes called a charge balance converter) introduces a sinc roll-off
\begin{equation}
H_{t_\mathrm{a}}=\frac{\sin(\pi f t_\mathrm{a})}{\pi f t_\mathrm{a}},
\label{eq:Hta}
\end{equation}
with a phase offset originating directly from the time delay introduced by one half of the aperture time
\begin{equation}
\Phi_{t_\mathrm{a}}=\pi f t_\mathrm{a},
\label{eq:Pta}
\end{equation}
where both \eqref{eq:Hta} and \eqref{eq:Pta} proved to be highly accurate\cite{Swerlein1991,LapuhBook} and precisely correctable for virtually any usable product of $f t_\mathrm{a}$\cite{LapuhBook,Lapuh2018a}.
It is worth noting that IADC resolution increases with an increasing aperture time, as the IADC has more time to integrate the input voltage and hence achieve higher resolving resolution. 
However, the final effective resolution obtained is almost always limited by the stability and noise of the whole system, but sometimes surprisingly with the internal numerical resolution of the digital firmware involved, as seems to be a case with 8588A at aperture times above \SI{2}{\second} or so.

Instead of using integrated analogue to digital converter, a sample-hold based analogue to digital converter, for example successive-approximation ADC with a sample-hold circuitry in the front-end, can be used to mimic the aperture time behaviour of the integrated analogue to digital converter by using averaging.
This was employed in Fluke 8588A DMM, where SAR ADC is always used to perform analogue to digital conversion for aperture times below \SI{100}{\micro\second} or effectively for sampling frequencies above \SI{4,7619}{\kilo\hertz}.
This way, SAR ADC can be used in a digitising mode to cover average aperture times up to \SI{3}{\milli\second.}.
The SAR ADC is operating at a constant sampling frequency ($f_\mathrm{sd}=\SI{5}{\mega\hertz}$ is a fixed value for Fluke 8588A) and the resulting conversion is calculated as an average over the programmed aperture time $t_\mathrm{ad}$
\begin{equation}
U_\mathrm{t_{ad}}=\frac{{\sum\limits_{n=1}^{Q} U_n}}{Q},
\label{eq:U_SAR_avg}
\end{equation}
where $U_n$ are individual readings from the SAR ADC conversion taken during that aperture time and $Q$ is the number of samples used for one average value calculation.
With a given sampling time $t_\mathrm{sd}$ and the number of samples taken $Q$, we get the resulting digital aperture time
\begin{equation}
t_\mathrm{ad}=Q t_\mathrm{sd},
\label{eq:t_ad}
\end{equation}
emulating the actual behaviour of the integrating (or charge balance) analogue to digital converter.
This process also introduces the frequency dependent roll-off \cite{Smith1997}, similar but not equal to \eqref{eq:Hta}
\begin{equation}
H_{t_\mathrm{ad}}=\frac{\sin(\pi Q \bar{f})}{Q \sin(\pi \bar{f})},
\label{eq:HtaSAR}
\end{equation}
where $\bar{f}$ is normalised signal frequency
\begin{equation}
\bar{f} = \frac{f}{f_\mathrm{sd}},
\label{eq:f_normalised}
\end{equation}
Eq. \eqref{eq:HtaSAR} can be rewritten into
\begin{equation}
H_{t_\mathrm{ad}}=\frac{\sin(\pi f t_\mathrm{ad})}{t_\mathrm{ad}/t_\mathrm{sd} \sin(\pi f t_\mathrm{sd})}.
\label{eq:Htad}
\end{equation}
which more closely resembles the $H_\mathrm{t_a}$ given by \eqref{eq:Hta}.
$H_{t_\mathrm{ad}}$ closely follows $H_{t_\mathrm{a}}$ for $\bar{f} {<\!\!\!<} 1$, but has a smaller attenuation of any frequency component approaching or being above $\bar{f}$ and a distinctive peak response at (no attenuation at exactly that frequencies) and around the positive integer multiples of $\bar{f}$.
\begin{figure}
\center
\includegraphics[scale = 1] {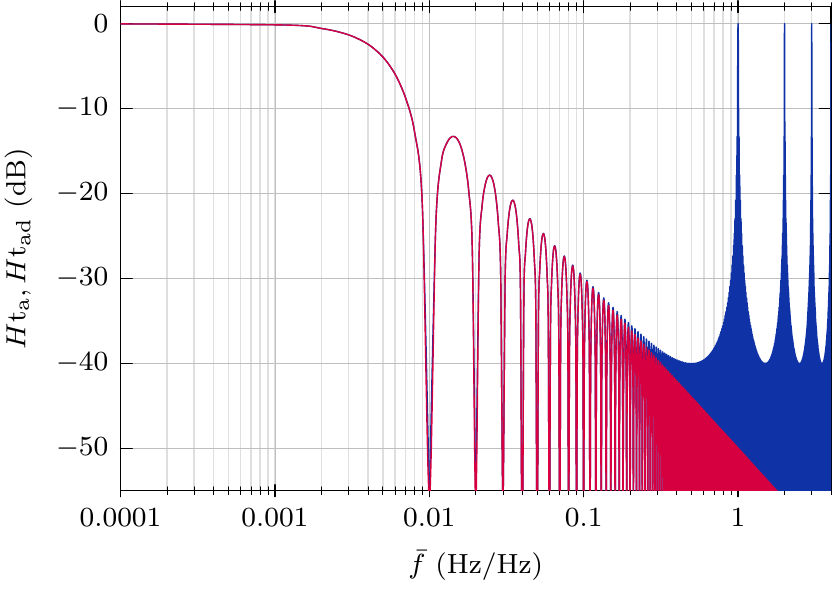}
\caption{IADC and averaging frequency response as a function of normalised frequency, where integration time $t_\mathrm{a}$ for IADC (red curve) equals the averaging time $t_\mathrm{ad}$ for SAR. $Q=100, t_\mathrm{a}=t_\mathrm{ad}=t_\mathrm{sd}/Q$.}
\label{fig:Sinc_IADC_SAR}
\end{figure}
Figure \ref{fig:Sinc_IADC_SAR} plots both $H_\mathrm{a}$ and $H_\mathrm{ad}$ for the case $t_\mathrm{a}=t_\mathrm{ad}$ for \num{100} samples being used for one averaging measurement result ($Q=100$).

\subsection{Input filter roll-off}\label{sec:Input filter roll-off}
For frequencies up to 20 kHz, the 3458A input filter amplitude response is often modelled with a single pole RC filter\cite{Swerlein1991}
\begin{equation}
    H_{1p}=\frac{1}{\sqrt{1+\left(\frac{f}{f_\mathrm{3dB}}\right)^2}},
    \label{eq:H1p}
\end{equation}
since its amplitude response was already shown to closely resemble the DMM response up to a few \si{\kilo\hertz}\cite{Gubler2010}, despite the input circuitry actually consists of two equal serially connected RC networks with \SI{1}{\percent} tolerance $\SI{5}{\kilo\ohm}$ resistors and \SI{2}{\percent} tolerance $\SI{82}{\pico\farad}$ capacitors.
The single pole RC filter model would implicate phase response
\begin{equation}
    \Phi_{1p}=\tan^{-1}\left(\frac{f}{f_\mathrm{3dB}}\right).
    \label{eq:P1p}
\end{equation}

\begin{figure}
\center
\includegraphics[scale = 1]{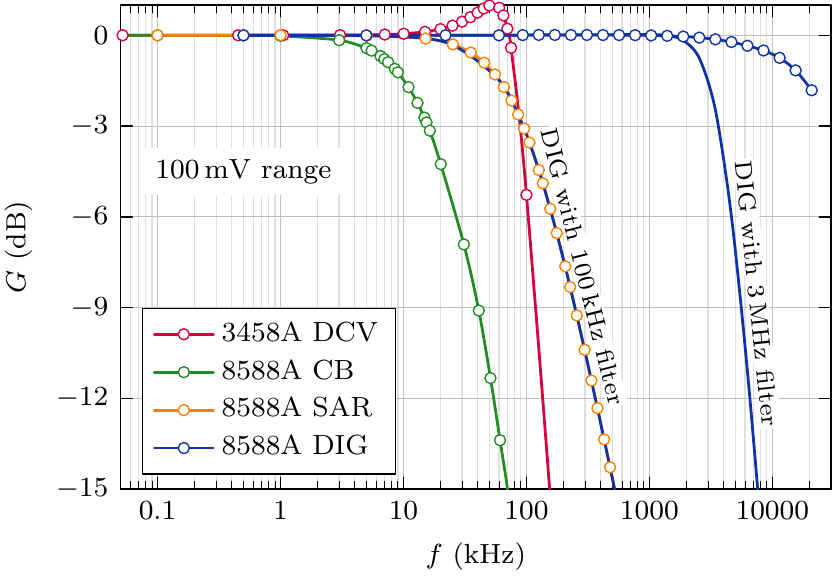}
\caption{\SI{100}{\milli\volt} range frequency response. The 8588A digitising mode with \SI{100}{\kilo\hertz} filter overlaps almost completely with the 8588A SAR response.}
\label{fig:Bandwidth 100mV}
\end{figure}

\begin{figure}
\center
\includegraphics[scale = 1] {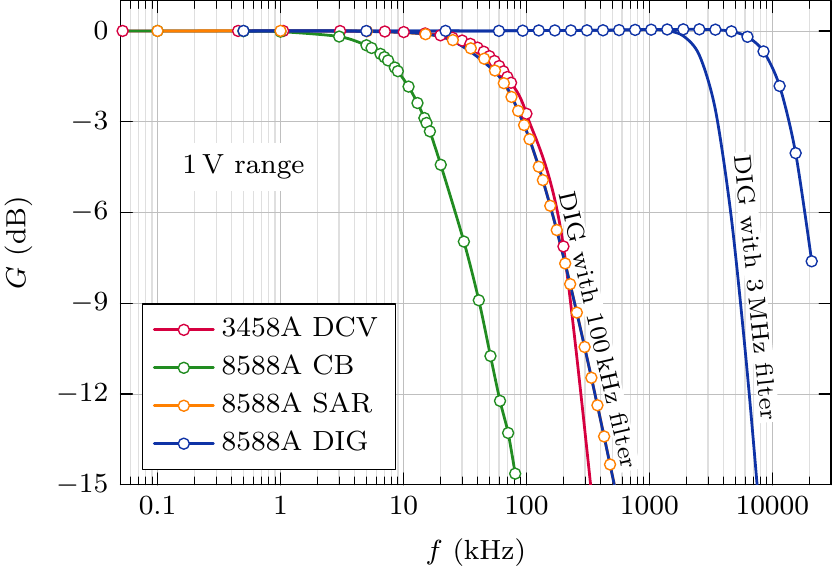}
\caption{\SI{1}{\volt} range frequency response.}
\label{fig:Bandwidth 1V}
\end{figure}

\begin{figure}
\center
\includegraphics[scale = 1] {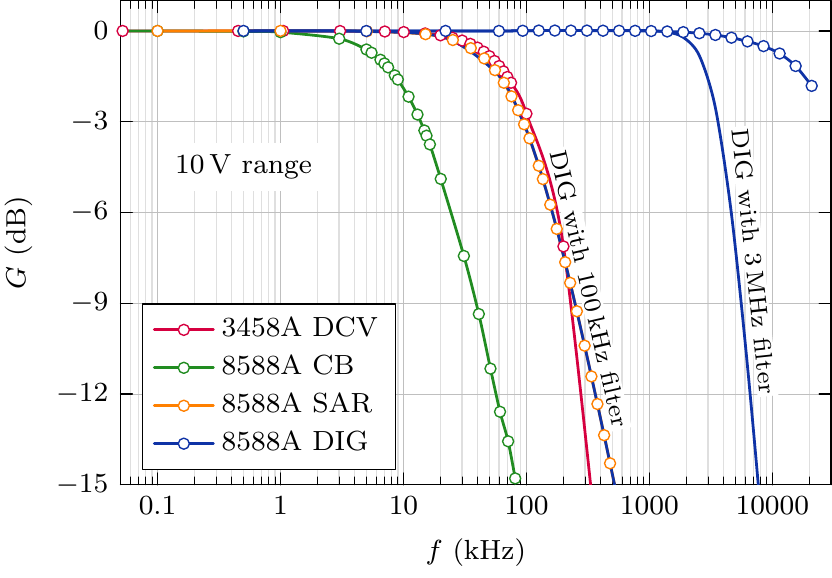}
\caption{\SI{10}{\volt} range frequency response.}
\label{fig:Bandwidth 10V}
\end{figure}

\begin{figure}
\center
\includegraphics[scale = 1] {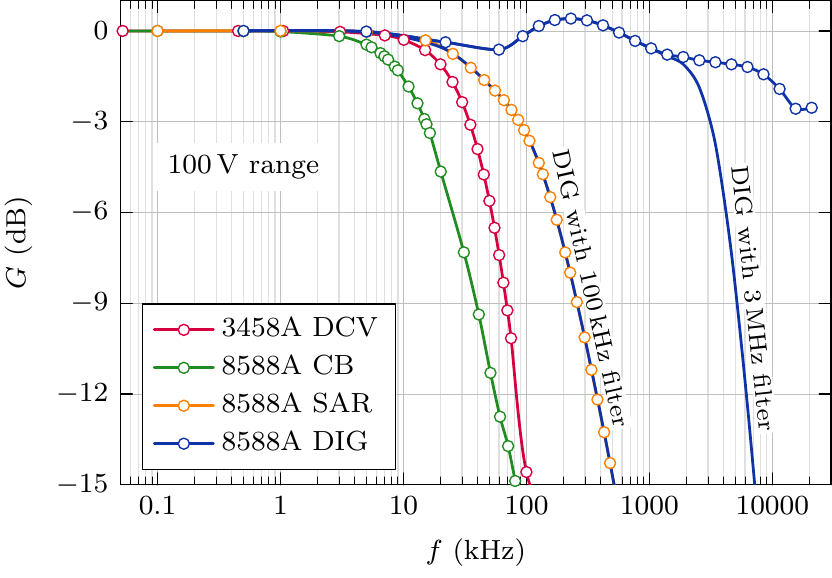}
\caption{\SI{100}{\volt} range frequency response.}
\label{fig:Bandwidth 100V}
\end{figure}

\begin{figure}
\center
\includegraphics[scale = 1] {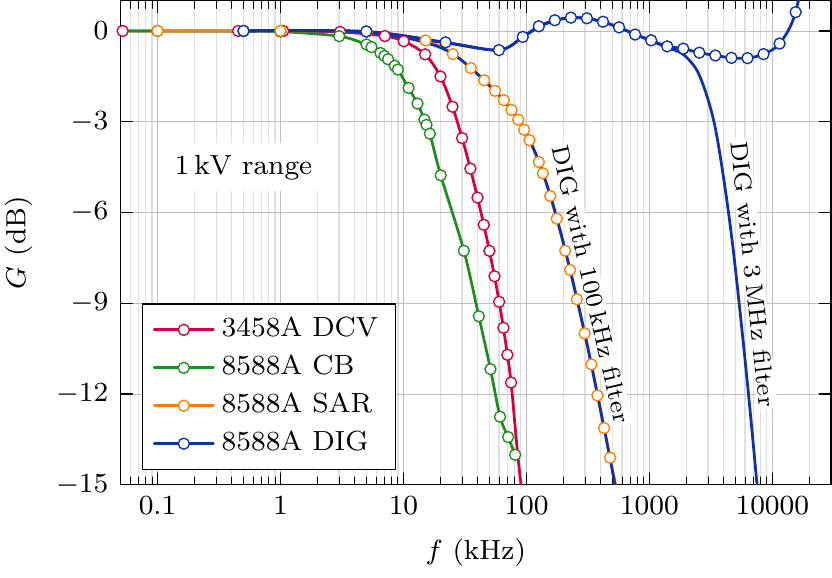}
\caption{\SI{1}{\kilo\volt} range frequency response.}
\label{fig:Bandwidth 1kV}
\end{figure}

The 8588A DMM nominal frequency response depends on the sampling mode and is collected in Table \ref{tab:Bandwidth8588A} for 10 V range.
\begin{table}
  \caption{Nominal frequency bandwidths for 8588A DMM voltage sampling modes. Values marked with * are not specified by a manufacturer.}
  \label{tab:Bandwidth8588A}
  \begin{center}
  \begin{tabular}{cccc}
  \toprule
  Mode  & Filter  & $f_\mathrm{3dB}$            & filter poles \\
  \midrule 
  CB    & -       & \SI{13.8}{\kilo\hertz}*      & \num{1} \\
  SAR	& -       & \SI{100}{\kilo\hertz}*       & \num{1}   \\
  DIG	& 100 kHz & \SI{100}{\kilo\hertz}       & \num{1} \\
  DIG	& 3 MHz   & \SI{3}{\mega\hertz}         & \num{4} \\
  DIG	& Off     & \SI{15}{\mega\hertz} - \SI{20}{\mega\hertz} & - \\
  \bottomrule
  \end{tabular}
  \end{center}
\end{table}
Measured frequency responses for all ranges and modes for both DMM types are plotted on Figures \ref{fig:Bandwidth 100mV} to \ref{fig:Bandwidth 1kV}.
All measurements were performed by applying a sine wave from arbitrary waveform synthesizer at selected frequencies and estimating a measured amplitude using an asynchronous PSFE sine wave amplitude estimation algorithm \cite{Lapuh2015b}. 
A \SI{120}{\mega\hertz} synthesizer bandwidth secured a sine wave level constant enough for the purpose over all applied frequencies up to \SI{20}{\mega\hertz}.
A full range voltages were set for \SI{100}{\milli\volt}, \SI{1}{\volt} and \SI{10}{\volt} ranges.
A \SI{7.07}{\volt} amplitude was used on \SI{100}{\volt} and \SI{1}{\kilo\volt} ranges, which still provided a reasonable measurement for DMM bandwidths up to \SI{100}{\kilo\hertz} or so, despite significantly decreased signal to noise ratio.
However, for higher bandwidths on \SI{100}{\volt} and \SI{1}{\kilo\volt} ranges, the measured results plotted on figures \ref{fig:Bandwidth 100V} and \ref{fig:Bandwidth 1kV} show deviation from a smooth low pass filter response.
This response might be attributed to the internal 8588A DMM voltage divider.

The measured responses show almost identical frequency bandwidth between 3458A DCV mode and 8588A SAR mode. The 8588A CB mode provides a very narrow bandwidth of \SI{13.8}{\kilo\hertz}, which already impacts a \SI{50}{\hertz} signal by \SI{-6.6}{\micro\volt/\volt}.

\IEEEpubidadjcol
\subsection{Distortions}\label{sec:Distortions}
\begin{figure}
\center
\includegraphics[scale = 0.95] {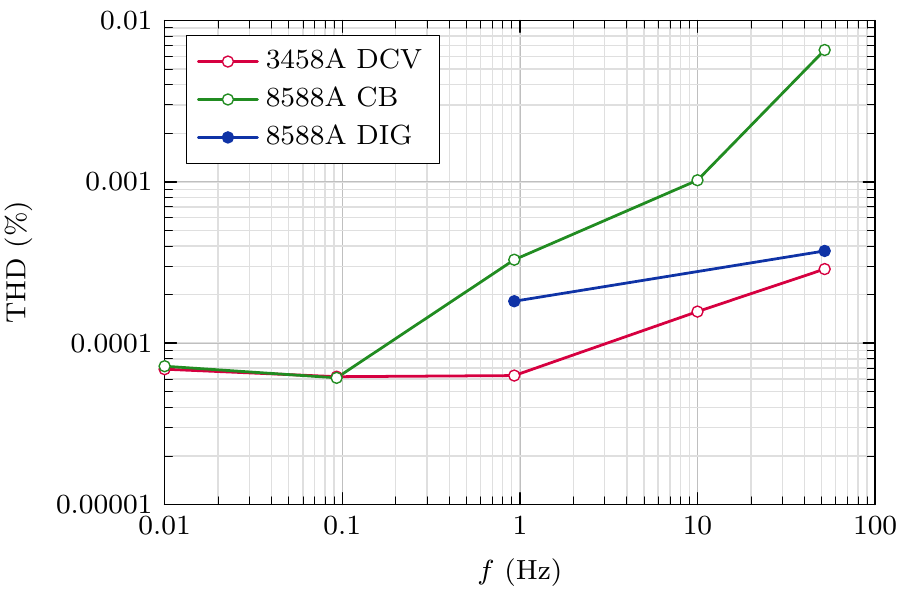}
\caption{8588A DMM charge balance ADC total harmonic distortions. 3458A DMM and 8588A DMM in DIG mode distortions are plotted measuring the same signal source with the same sampling parameters.}
\label{fig:THDvsFREQ_IADC}
\end{figure}

While performing tests in 8588A CB mode using a precision \v{C}MI SWG03 \cite{Kucera2020} voltage source and observing frequency spectrum obtained, two specific results were noted.
First was an elevated noise floor at signal frequencies above around \SI{10}{\hertz} due to the internal 8588A \SI{100}{\nano\second} time jitter (see Figure \ref{fig:Jitter noise CB}.
Second were a persistent third and fifth harmonic components, being clearly visible at signal frequencies above \SI{1}{\hertz}.
Figure \ref{fig:THDvsFREQ_IADC} plots THD as a function of input signal frequency. 
As the THD of the \v{C}MI SWG03 source was not estimated separately, two DMMs were connected to the same signal in parallel. 
Their measured THD is plotted as 8588A (DIG) and 3458A (DCV) and demonstrates the source THD, being equal to 3458A (DCV) plotted curve or better.
Table \ref{tab:THDparameters} lists sampling parameters that were used for all three cases at each test frequency.
Measurements were performed in a master-slave configuration and at each frequency two measurements were performed with the 8588A DMM in CB mode operating as a master and as a slave. 
Apart from reducing the SNR of the slave DMM when 8588A DMM in CB mode was used as a master and consequently masking the harmonics of that slave SNR at frequencies above \SI{10}{\hertz}, THD results of the 8588A DMM in CB mode were not differing significantly between master and slave mode.
A multi-frequency sine fitting algorithm \cite{Vandersteen1997} was used to estimate THD.
\begin{table}
  \caption{DMM sampling parameters used to perform THD measurements. Signal RMS amplitude was \SI{7}{\volt}.}
  \label{tab:THDparameters}
  \begin{center}
  \begin{tabular}{cccc}
  \toprule
  $f$  & $t_\mathrm{s}$  & $t_\mathrm{a}$ & $N$ \\
  \midrule 
  \SI{10}{\milli\hertz}  & \SI{1}{\second}          & \SI{980}{\milli\second} & \num{1000} \\
  \SI{93}{\milli\hertz} & \SI{90}{\milli\second}   & \SI{89}{\milli\second} & \num{10000} \\
  \SI{930}{\milli\hertz}  & \SI{9}{\milli\second}    & \SI{1}{\milli\second} & \num{10000} \\
  \SI{10}{\hertz}    & \SI{1}{\milli\second}    & \SI{890}{\micro\second} & \num{37888} \\
  \SI{52}{\hertz}    & \SI{1}{\milli\second}    & \SI{890}{\micro\second} & \num{40000} \\
  \bottomrule
  \end{tabular}
  \end{center}
\end{table}
The 8588A operating in CB mode shows an elevated harmonic distortions at and above \SI{1}{\Hz}.
\begin{figure}
\center
\includegraphics[scale = 0.95] {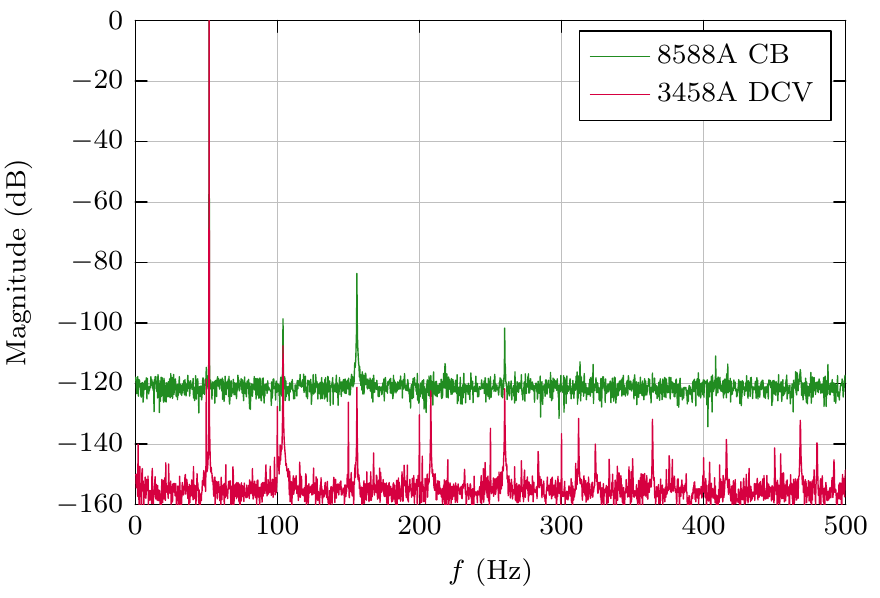}
\caption{8588A DMM CB mode and 3458A DCV mode spectrum while sampling \SI{52}{\hertz} signal with $t_\mathrm{s} = \SI{1}{\milli\second}$, $t_\mathrm{a} = \SI{890}{\micro\second}$, $N = 4000$ and $M = 4$. Signal source was \v{C}MI SWG03 and signal amplitude was \SI{7}{\volt}.},
\label{fig:Spectrum_SAMPLER}
\end{figure}
Figure \ref{fig:Spectrum_SAMPLER} plots a spectrum of the same signal obtained by 8588A and 3458A.
A multi-frequency sine fitting algorithm was used to first remove signal components from the sampled data. 
FFT was used to retrieve the spectrum of this residual and harmonic components removed from the signal were placed directly into the spectrum results as signal peaks.
Clearly, 8588A result is masked by the noise, but it also shows much stronger second, third and fifth harmonics, which are obviously not present in the signal.
Resulting spectrum using 3458A clearly shows line frequency (\SI{50}{\hertz} components up to the fifth harmonic, which is possible due to a very good \SI{-151}{\decibel} noise floor achieved at the selected sampling parameters.

As it is expected for both 3458A and 8588A DMMs to be linear far below \SI{1}{\micro\volt/\volt} at DC input signal, this result shows that 8588A CB mode dynamic linearity is not following this expectation.
However, 8588A DIG mode performs better and shows no signs of increased THD at any frequency. 
At \SI{1}{\kilo\hertz}, it was measured at \SI{0,0025}{\percent} using the \v{C}MI SWG03 source.
At \SI{110}{\kilo\hertz}, it was measured at \SI{0,018}{\percent} using the Keysight 33622A source.
However, these results present the combined THD of both the source and the sampler and can only be used as a potential performance indication.

\section{DMM Phase Response}\label{sec:DMM Phase Response}
Phase response of a single DMM is usually not of interest and is also never specified by the DMM manufacturer.
However, numerous applications for measuring electrical power \cite{LapuhBook} make direct use of phase measurement using two sampling DMMs.
To enable phase measurement, two DMM should be synchronised, either by a common trigger signal or operating in a master-slave fashion.
To remove residual phase measurement errors due to various time delays of such setup, a known phase should be measured to provide for a reference point.
An ideal \SI{0}{\radian} phase reference is a single sine wave signal at a given frequency, which is split to both DMM inputs using equal length cables.
By interchanging the cables between two DMMs and comparing time delay results, the cable and DMM mismatch can be minimised.

While 3458A DMM has only an external trigger input to facilitate synchronisation, 8588A DMM adds a \SI{1}{\mega\hertz} or \SI{10}{\mega\hertz} external clock input, which greatly simplifies synchronous measurements using one or more 8588A. 
Unfortunately, it does not have any clock output.

To fully evaluate phase performance of the samplers, one would need to have an ideal phase source, which of course is not available.
Instead, an arbitrary waveform generator Keysight 33622A and a custom made \v{C}MI SWG03 high stability DDS source were used. 
Their performance directly influences results presented so the results should be viewed as a superposition of source and sampler results. 
A few different configurations were used in measurement with an attempt to discriminate which instrument is affecting a non-ideal result, with some success.

There are different phase characteristics that can be attributed to a pair of DMMs sampling a two channel signal.
One is a phase linearity over the $2\pi$ phase angle at single frequency and all sampling parameters (like sampling time, aperture time, range, signal level) being fixed.
The other is a phase offset due to a trigger timing misalignment and different input circuitry phase delays, again at these same sampling parameters.
Finally, these two characteristics can be measured at different frequencies to obtain their frequency dependent response.
The phase results presented here were not performed thoroughly to give a full picture for either of them as the equipment used was not available for all those measurements.
However, I hope they provide a sneak peek into what is possible to achieve in terms of phase accuracy over a reasonably large frequency bandwidth.

For all phase results presented in this paper, \SI{7}{\volt} signals were used on \SI{10}{\volt} DMM's range.

\subsection{Master-Slave synchronisation}
\v{C}MI SWG03 source was used for these measurements.
The settings given in Table \ref{tab:PhaseMS} were used.
The 3458A DMM was operating in a DCV mode, while 8588A DMM was operating in DCV mode only for \SI{52}{\hertz} signal frequency.
\num{100000} samples were taken for each phase measurement and \num{10} repetitions were performed at each phase setting.
\begin{table}
  \caption{Setting for master-slave phase measurements and results for zero phase measurement. 3458A DMM as master and 8588A as slave were used.}
  \label{tab:PhaseMS}
  \begin{center}
  \begin{tabular}{cccccc}
  \toprule
  $f$  & $t_\mathrm{a}$ & $t_\mathrm{s}$ & $\Delta t_\mathrm{0}$ & $\Delta \phi_\mathrm{0}$ & $\Delta \phi_\mathrm{dev}$\\
  \midrule 
  \SI{52}{\hertz} & \SI{195}{\micro\second} & \SI{220}{\micro\second} & \SI{1.647}{\micro\second} & \SI{0.22}{\micro\radian} & \SI{0.70}{\micro\radian}\\
  \SI{1}{\kilo\hertz} & \SI{1.4}{\micro\second} & \SI{10}{\micro\second} & \SI{1.668}{\micro\second} & \SI{2.4}{\micro\radian} & \SI{1.5}{\micro\radian}\\
  \SI{19}{\kilo\hertz} & \SI{1.4}{\micro\second} & \SI{10}{\micro\second} & \SI{1.649}{\micro\second} & \SI{293}{\micro\radian} & \SI{6.8}{\micro\radian}\\
  \bottomrule
  \end{tabular}
  \end{center}
\end{table}
As the 3458A DMM was used as a master DMM for final results, the synchronisation was not possible and thus a PSFE algorithm \cite{Lapuh2015b} was used to retrieve phase results.

\begin{figure}
\center
\includegraphics[scale = 1] {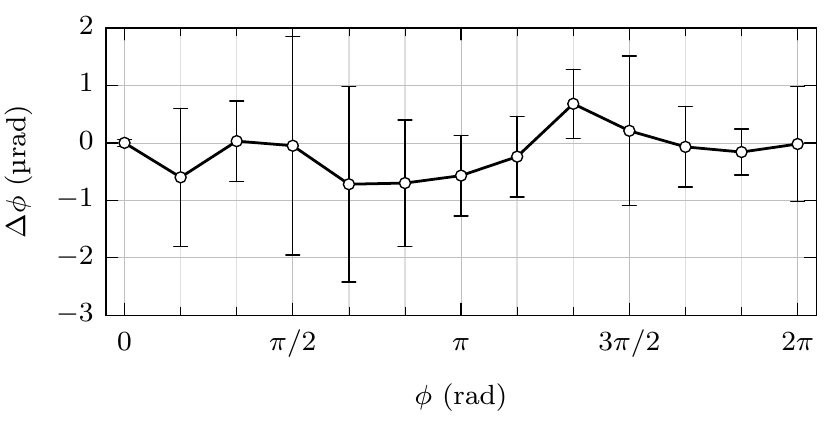}
\caption{Phase response at \SI{52}{\hertz} for 3458A DMM (master) and 8588A DMM (slave) configuration.}
\label{fig:Phase_52Hz_FK_S}
\end{figure}

\begin{figure}
\center
\includegraphics[scale = 1] {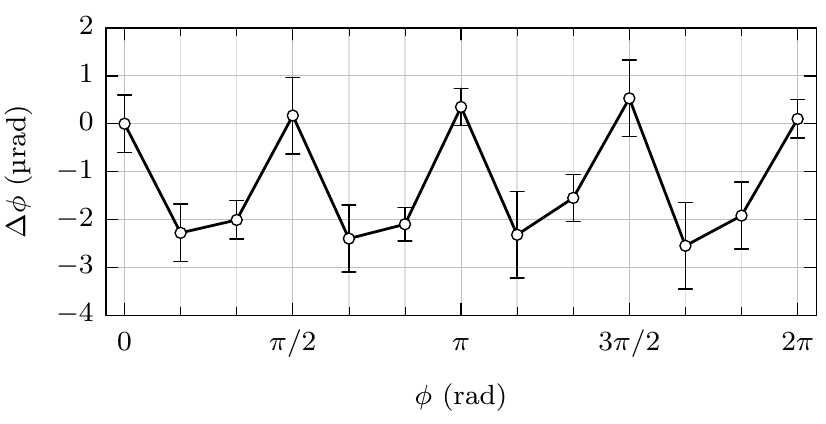}
\caption{Phase response at \SI{1}{\kilo\hertz} for 3458A DMM (master) and 8588A DMM (slave) configuration on \SI{10}{\volt} range.}
\label{fig:Phase_1kHz_FK_S}
\end{figure}

\begin{figure}
\center
\includegraphics[scale = 1] {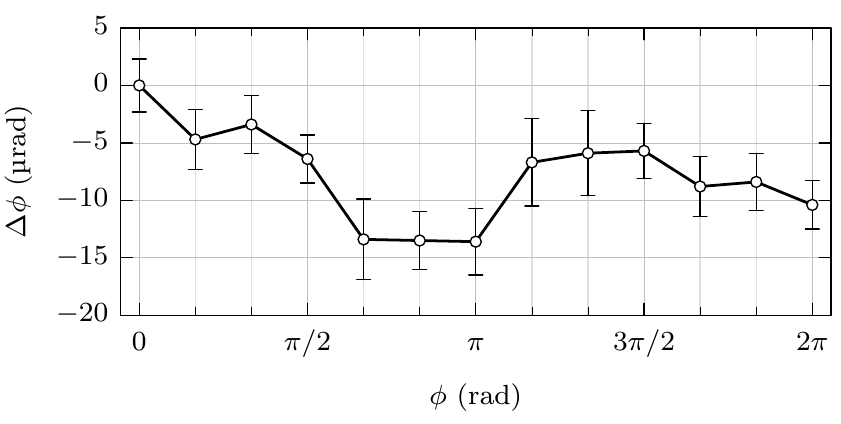}
\caption{Phase response at \SI{19}{\kilo\hertz} for 3458A DMM (master) and 8588A DMM (slave) configuration on \SI{10}{\volt} range.}
\label{fig:Phase_19kHz_FK_S}
\end{figure}

The following steps were performed:
\begin{enumerate}
    \item A single source channel was connected in parallel to both DMMs using a T-piece and a same length cables. By measuring residual phase, $\Delta t_\mathrm{0}$ was determined at each frequency, indicating a cumulative time delay between master and slave DMM.
    \item Source phase was programmed to \SI{0}{\degree}, both channels were connected to corresponding DMMs and the phase was measured again, taking into account a residual phase obtained in the previous step ($\Delta t_\mathrm{0}$). This resulted in a zero phase difference $\Delta \phi_\mathrm{0}$.
    \item Source phase between two channels was increased in steps of \SI{30}{\degree} to obtain phase linearity results within $\pm\Delta \phi_\mathrm{dev}$ as shown on figures \ref{fig:Phase_52Hz_FK_S}, \ref{fig:Phase_1kHz_FK_S} and \ref{fig:Phase_19kHz_FK_S}. These results are normalised to start from \SI{0}{\radian}. Vertical bars represent standard deviations obained from ten repetitions at each frequency and phase setting.
\end{enumerate}

\begin{figure}
\center
\includegraphics[scale = 1] {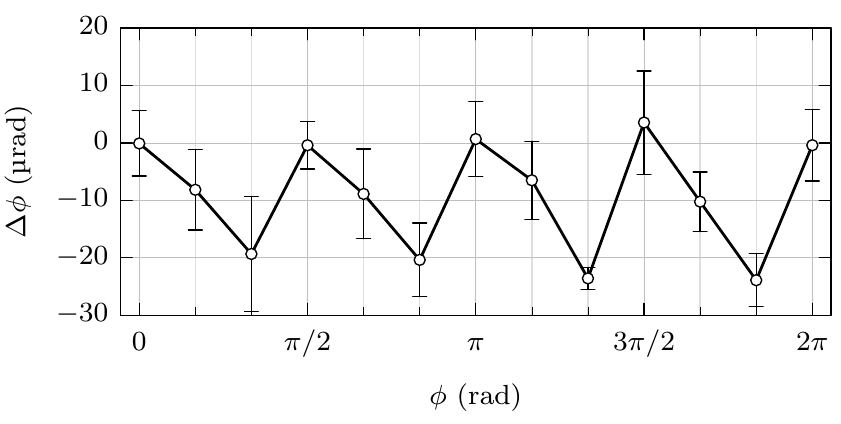}
\caption{Phase response at \SI{1}{\kilo\hertz} for 8588A DMM (master) and 3458A DMM (slave) configuration on \SI{10}{\volt} range.}
\label{fig:Phase_1kHz_FK_M}
\end{figure}

A question will certainly appear as to why the 3458A DMM was used as a master, rendering the clock synchronisation impossible.
This decision was made based on preliminary experiments, which showed that this configuration is far superior form using 8588A DMM as a master.
This is demonstrated on Figure \ref{fig:Phase_1kHz_FK_M}, where 8588A served as a master with all other settings remaining exactly the same as for the results shown on Figure \ref{fig:Phase_1kHz_FK_S}, in this case for $f=\SI{1}{\kilo\hertz}$.
Additionally, the same difference was observed at other signal frequencies.

No further research was done to find the reason for such behaviour and therefore no explanation can be provided at this point.
Unfortunately, no phase measurements were performed with \v{C}MI SWG03 source and two 8588A as when an operational source arrived, one 8588A DMM needed to be returned.

\begin{figure}
\center
\includegraphics[scale = 0.98] {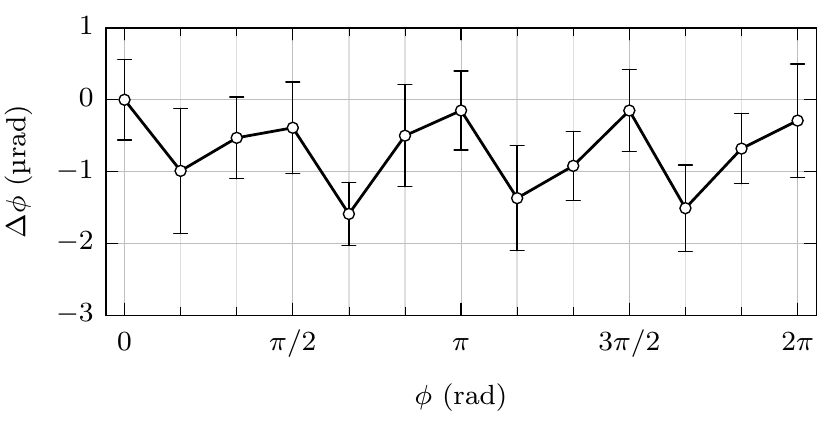}
\caption{Phase response at \SI{1}{\kilo\hertz} for 3458A DMM (master) and 8588A DMM (slave) configuration on \SI{10}{\volt} range. The source had a special function DAC aligned being enabled.}
\label{fig:Phase_1kHz_FK_S DAC aligned corrected}
\end{figure}

Another example is provided here for additional info only. 
The \v{C}MI SWG03 source software platform offers a DAC alignment mode.
This alignment secures a generation of an indentical signal even when changing its phase.
This is achieved by programming DACs with exactly the same words, regardless of the phase setting.
The price paid is that the phase setting resolution is reduced, but the actual phase setting is always reported and could be used as a reference.
The results from this experiment are presented on Figure \ref{fig:Phase_1kHz_FK_S DAC aligned corrected}. 
Comparing this with the results on Figure \ref{fig:Phase_1kHz_FK_M}, we can observe slightly lower standard deviations and slightly better measured phase linearity.
There is also a pattern seen on these results, repeating on every $\pi / 2$ \si{\radian} phase change, that also appears on Figures \ref{fig:Phase_1kHz_FK_S} and \ref{fig:Phase_1kHz_FK_M}.
This could be related to an effect of high frequency glitches coming from source DAC's steps, noting that DAC steps and glitches are hardly filtered out completely. 

\subsection{Common clock synchronisation}
For synchronous measurements, two 8588A DMMs were used with a common \SI{10}{\mega\hertz} clock supplied from Keysight 33622A arbitrary waveform synthesizer.
They were still configured in a master-slave fashion using Trig Out and Trig In connectors.
Both DMMs were configured to operate in Digitizing mode, using a SAR DAC.
\num{20000} samples were taken for each phase measurement and \num{10} repetitions were performed at each phase setting.
The setting given in Table \ref{tab:PhaseFF} were used.
\begin{table}
  \caption{Setting for clock synchronised master-slave phase measurements and results for zero phase measurement. Two 8588A DMM were used.}
  \label{tab:PhaseFF}
  \begin{center}
  \begin{tabular}{cccccc}
  \toprule
  $f$  & $t_\mathrm{a}$ & $t_\mathrm{s}$ & $\Delta t_\mathrm{0}$ & $\Delta \phi_\mathrm{0}$ & $\Delta \phi_\mathrm{dev}$\\
  \midrule 
  \SI{1.107}{\kilo\hertz} & \SI{48}{\micro\second} & \SI{50}{\micro\second} & \SI{104.68}{\nano\second} & \SI{-42.7}{\micro\radian} & \SI{29}{\micro\radian}\\
  \SI{11.075}{\kilo\hertz} & \SI{18}{\micro\second} & \SI{20}{\micro\second} & \SI{104.73}{\nano\second} & \SI{-6.7}{\micro\radian} & \SI{36}{\micro\radian}\\
  \SI{110.75}{\kilo\hertz} & \SI{0.6}{\micro\second} & \SI{1}{\micro\second} & \SI{104.80}{\nano\second} & \SI{49.7}{\micro\radian} & \SI{15}{\micro\radian}\\
  \SI{998.7}{\kilo\hertz} & \SI{0.1}{\micro\second} & \SI{0.4}{\micro\second} & \SI{104.82}{\nano\second} & \SI{1375}{\micro\radian} & \SI{38}{\micro\radian}\\
  \bottomrule
  \end{tabular}
  \end{center}
\end{table}

\begin{figure}
\center
\includegraphics[scale = 1] {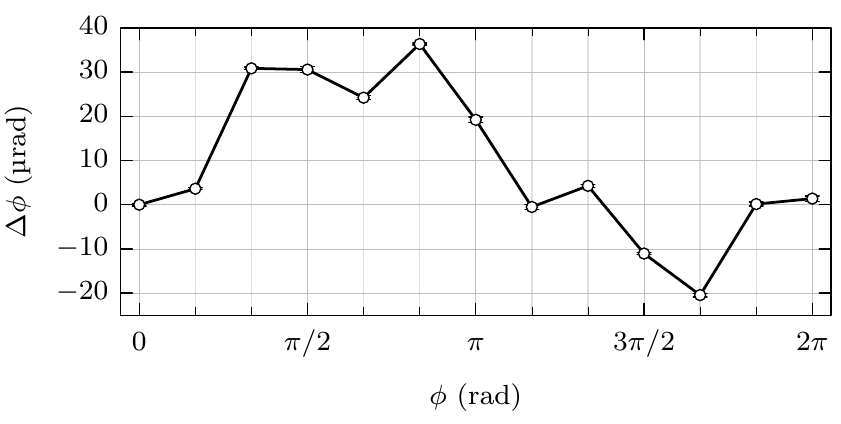}
\caption{Phase response at \SI{1}{\kilo\hertz} for two 8588A DMM using a common external \SI{10}{\mega\hertz} clock and internal TIMER for sampling instants.}
\label{fig:Phase_1kHz_FFdig}
\end{figure}

\begin{figure}
\center
\includegraphics[scale = 1] {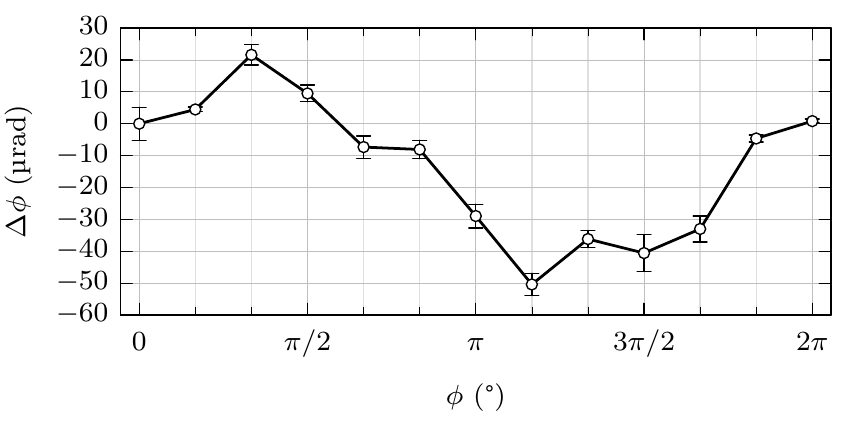}
\caption{Phase response at \SI{11}{\kilo\hertz} for two 8588A DMM using a common external \SI{10}{\mega\hertz} clock and internal TIMER for sampling instants.}
\label{fig:Phase_11kHz_FFdig}
\end{figure}

\begin{figure}
\center
\includegraphics[scale = 1] {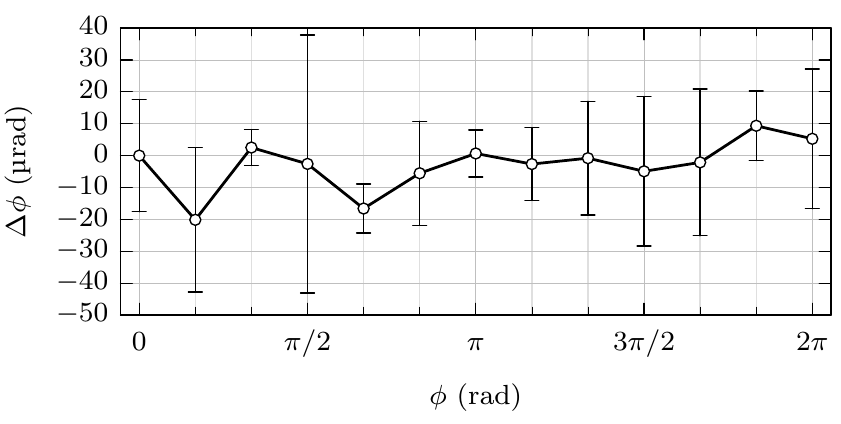}
\caption{Phase response at \SI{110}{\kilo\hertz} for two 8588A DMM using a common external \SI{10}{\mega\hertz} clock and internal TIMER for sampling instants.}
\label{fig:Phase_110kHz_FFdig}
\end{figure}

\begin{figure}
\center
\includegraphics[scale = 1] {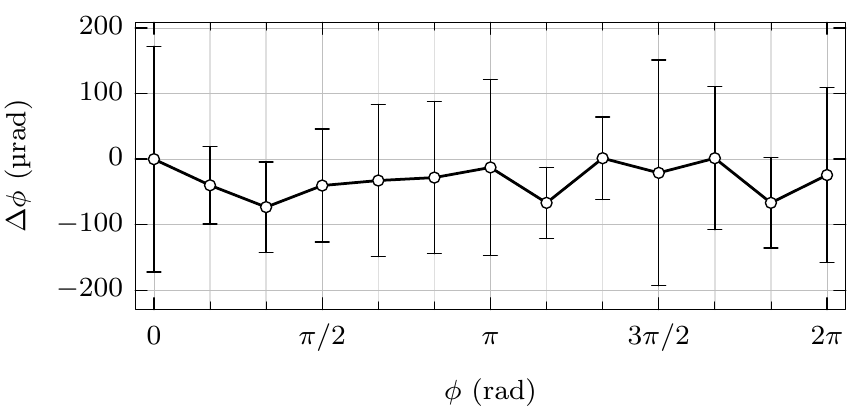}
\caption{Phase response at \SI{1}{\mega\hertz} for two 8588A DMM using a common external \SI{10}{\mega\hertz} clock and internal TIMER for sampling instants. Measurements performed on \SI{10}{\volt} range.}
\label{fig:Phase_1MHz_FFdig}
\end{figure}

As the sampling was synchronous, all results were obtained using FFT.
The same procedure was used as for the master-slave configuration.
Results are presented on figures \ref{fig:Phase_1kHz_FFdig}, \ref{fig:Phase_11kHz_FFdig}, \ref{fig:Phase_110kHz_FFdig} and \ref{fig:Phase_1MHz_FFdig}.

While these results are quite promising, they do not show whether source or DMMs are responsible for non-linearity at \SI{1}{\kilo\hertz} and \SI{11}{\kilo\hertz}, where the standard deviations are much smaller than this non-linearity. 

\section{Discussion}\label{sec:Discussion}
The 8588A DMM is a first serious challenge for 3458A DMM dominance in state of the art sampling accuracy at low frequencies.
It sports a huge internal memory, all modern interface options and a clock input for timing synchronisation.
Its functionality limits a direct use of integrating ADC for relatively slow sampling ($f_\mathrm{s}<=\SI{4.7619}{\kilo\hertz}$).
While its CB path design incorporates a zero drift amplifier technology which effectively pushes $1/f$ noise corner to extremely low frequencies, the price paid is a much elevated white noise floor as well as quite inferior jitter performance, both as compared to 3458A DMM. 
This makes the 8588A DMM extremely suitable for measuring DC (and extremely low frequencies) signals. 
However, the available frequency band in this mode is limited and combined with elevated white noise, timing jitter and harmonic distortions, the 8588A overall sampling performance is inferior already for sampling \SI{50}{\hertz} frequency signals.

For sampling faster signals, the 8588A DMM relies on averaging \SI{18}{bit} SAR ADC technology for sampling frequencies up to \SI{5}{\mega\hertz}.
The performance of this digitising mode looks excellent, but it no longer holds the same DC accuracy, $1/f$ noise corner and stability demonstrated by CB path.
Additionally, it is not clear if this this mode surpasses the commercial SAR ADC technology available and how it will cope with a very rapid developments in this field. 
It is unlikely it would achieve and maintain the same dominance as 3458A IADC did over the last 30 years.

\section{Further work}
It would be interesting to compare the actual DC linearity of 3458A and 8588A DMMs with a DC Josephson source.
Additionally, phase performance should be measured more accurately, as it is quite possible that the results presented here were influenced more by a source then by a DMMs.
Certainty, further investigations of this new DMM will reveal its performance in specific areas and applications.

\section*{Acknowledgment}
The author would like to thank Fluke Corporation and Micom to lend two 8588A DMMs for a substantial amount of time, Matja\v{z} Lindi\v{c} from SIQ for lending Keysight 3458A and 33622A instruments and Jan Ku\v{c}era from \v{C}MI for lending their SWG03 source. A special thanks goes to Bo\v{s}tjan Volj\v{c}, being always ready to help with virtually everything I needed.

\ifCLASSOPTIONcaptionsoff
  \newpage
\fi

\bibliographystyle{IEEEtran} 
\bibliography{mybib}

\end{document}